# PUMP AND PROBE NONLINEAR PROCESSES: NEW MODIFIED SUM RULES FROM A SIMPLE OSCILLATOR MODEL

F. Bassani and V. Lucarini

Scuola Normale Superiore, 56100 Pisa

and

Istituto Nazionale di Fisica della Materia, Italy

## ABSTRACT

The nonlinear oscillator model is useful to basically understand the most important properties of nonlinear optical processes. It has been shown to give the correct asymptotic behaviour and to provide the general features of harmonic generation to all orders, in particular dispersion relations and sum rules. We investigate the properties of Pump and Probe processes using this model, and study those cases where general theorems based on the holomorphic character of the Kubo response functions cannot be applied. We show that it is possible to derive new sum rules and new Kramers-Krönig relations for the two lowest momenta of the real and of the imaginary part of the third order susceptibility and that new specific contributions become relevant as the intensity of the probe increases. Since the analitic properties of the susceptibility functions depend only upon the time causality of the system we are confident that these results are not model dependent and therefore have a general validity, provided one substitutes to the equilibrium values of the potential derivatives the density matrix expectation values of the corresponding operators.

P.A.C.S. numbers: 42.65.An, 42.65.Dr, 42.65.Sf, 78.20.Bh, 78.20.Ci

# I. INTRODUCTION

The theoretical and experimental investigation of systems responding to Pump and Probe optical beams is an important branch of research in solid state physics [1,2,3,4,5,6]. The most important nonlinear effect is the modification of the optical response near each resonance, which is often called the dynamical Stark effect [7,8,9,10,11]. Other new effects are however present, such as the two-photons absorption [12,13,14,15] and the stimulated Raman scattering [16,17,18,19,20]. To completely interpret the above phenomena detailed calculations are necessary [21,22,23,24,25] but in general they are very difficult to perform with the required accuracy for realistic systems. For this reason general properties of the nonlinear response functions are very useful for the interpretation of experimental data and the testing of appropriate models [26,27,28]. In particular, sum rules and Kramers-Krönig (K.K.) relations are expected to be of great help also in nonlinear optics, because they only depend on general properties, such as time causality, and must be verified by any physical system.

Sum rules and K.K. relations have been derived for those specific nonlinear processes represented by the susceptibility $\chi^{(n)}(\omega_1, \omega_2, -\omega_2, ..., \omega_2, -\omega_2)$, which is the most relevant for weak probes (whose frequency is $\omega_1$) since it is proportional to the $(n-1)/2^{th}$ power of the intensity of the pump beam. In fact it has the property of being holomorphic in the upper complex half plane of the variable $\omega_1$ thanks to Scandolo's theorem [29].

When we consider nonlinear effects also due to the probe beam the previous treatment is not valid anymore and new poles in the $n^{th}$ order susceptibility functions appear in the upper complex plane. Scandolo's theorem in this case is not verified so that nothing can be said about K.K. relations and sum rules from the previous analyses. This is clearly explained in the book by Peiponen et al. [30]

The main purpose of the present work is to study in detail the third order susceptibility $\chi^{(3)}(\omega_1)$ with a probe beam and a pump beam, and to derive general properties such as K.K. relations and sum rules. Since such general properties do not depend on the microscopic decription of the nonlinear system, we adopt a simple anharmonic oscillator model along the line described by Bloembergen [1] and Garrett [31,32] and previously adopted for the studty of harmonic generation [33] and for general studies of sum rules in special cases [34,35,36]. Such a model allows a derivation of the most important features, considering all nonlinearities and including also the nonlinear effects arising from the probe beam. We will show that new K.K. relations and modified sum rules can be obtained, taking into account optical rectification and considering the poles in the upper complex plane, which originate from the probe nonlinearities.

From previous experience with harmonic generation processes [33] we feel confident that the results here obtained can be extended to the quantum mechanical analysis by substituting the expectation values of the potential derivatives to their values at equilibrium used in the classical description.

In section II we describe our model and give formulas for $\chi^{(3)}(\omega_1)$, while the higher orders expressions are given in the Appendix. We analyze the structure of the third order susceptibility and we show that the holomorphic and non holomorphic contributions can be separated and their respective properties can be considered. In section III. we derive our modified sum rules. In section IV we derive the K.K. relations and discuss their connection with the sum rules. In section V we present our conclusions.

## II. THE NONLINEAR OSCILLATOR MODEL AND THE THIRD ORDER SUSCEPTIBILITY WITH NON HOLOMORPHIC CONTRIBUTIONS

We adopt the Lorentz oscillator model with a general potential energy mV(x) containing anharmonic terms, introduce a damping $\gamma$ and consider two incident optical beams, one used as a probe of frequency $\omega_1$, and one as a pump beam of frequency $\omega_2$:

$$E(t) = E_1 e^{-i\omega_1 t} + E_2 e^{-i\omega_2 t} + c.c. \quad , \tag{1}$$

where c.c. indicates the complex conjugates of the previous terms.

Expanding the potential around the equilibrium position we obtain the following equation of motion:

$$\ddot{x} + \gamma\dot{x} + \omega_0^2 x + \sum_{n=3}^{\infty}\left[\frac{\partial^n V(x)}{\partial x^n}\right]_0 \frac{x^{n-1}}{(n-1)!} = \frac{eE(t)}{m} \quad (2)$$

The solution of this equation can be obtained with an iterative procedure as described in ref. [1,2] and recently reported in detail in the book by Peiponen et al. [30].

We obtain for the linear contribution:

$$P^{(1)}(\omega_j) \equiv E_j \chi^{(1)}(\omega_j), \qquad j=1,2 \quad (3a)$$

with the usual :

$$\chi^{(1)}(\omega_j) = \frac{e^2 N/m}{D(\omega_j)} \quad (3b)$$

where :

$$D(\omega_j) = \omega_0^2 - \omega_j^2 - i\gamma\omega_j \quad . \quad (3c)$$

A general expression of the nonlinear Polarization $P^{(n)}(r\omega_1 + s\omega_2)$ is given in the Appendix A, where also a formal analysis of the properties of such functions in the complex $\omega_1$ plane is given. We here concentrate on the third order susceptibility, and wish to prove in this particular case that general results like K.K. relations and sum rules can be extended to include the contributions which are nonholomorphic in the upper complex $\omega_1$ plane, provided the poles are explicitely considered.

The expression of the $\chi^{(3)}(\omega_1)$ can be obtained in the usual way by successive iterations [1,30,33] and also follows from the general result given in the Appendix A substituting n=3, r=1 and s=0 in (A1). We obtain:

$$\chi^{(3)}(\omega_1) = \chi^{(3)}(\omega_1;\omega_1,-\omega_2,\omega_2) + \chi^{(3)}(\omega_1;\omega_1,0(-\omega_2,\omega_2)) +$$

$$+ \chi^{(3)}(\omega_1;\omega_2,\omega_1-\omega_2) + \chi^{(3)}(\omega_1;-\omega_2,\omega_1+\omega_2)$$

$$+ \chi^{(3)}(\omega_1;\omega_1,\omega_1,-\omega_1) + \chi^{(3)}(\omega_1;\omega_1,0(\omega_1,-\omega_1)) + \chi^{(3)}(\omega_1;-\omega_1,2\omega_1) \quad , \quad (4a)$$

where :

$$\chi^{(3)}(\omega_1;\omega_1,-\omega_2,\omega_2) = -\left[\frac{\partial^4 V(x)}{\partial x^4}\right]_0 \frac{|E_2|^2 e^4 N/m^3}{D(\omega_1)^2 D(\omega_2) D(-\omega_2)} \quad , \quad (4b)$$

$$\chi^{(3)}(\omega_1;\omega_1,0(\omega_2,-\omega_2)) = \left[\frac{\partial^3 V(x)}{\partial x^3}\right]_0^2 \frac{|E_2|^2 e^4 N/m^3}{\omega_0^2 D(\omega_1)^2 D(\omega_2) D(-\omega_2)} \quad , \quad (4c)$$

$$\chi^{(3)}(\omega_1;\pm\omega_2,\omega_1 \mp \omega_2) = \left[\frac{\partial^3 V(x)}{\partial x^3}\right]_0^2 \frac{|E_2|^2 e^4 N/m^3}{D(\omega_1)^2 D(\omega_2) D(-\omega_2) D(\omega_1 \mp \omega_2)} \quad , \quad (4d\text{-}4e)$$

$$\chi^{(3)}(\omega_1;\omega_1,-\omega_1,\omega_1) = -\frac{1}{2}\left[\frac{\partial^4 V(x)}{\partial x^4}\right]_0 \frac{|E_1|^2 e^4 N/m^3}{D(\omega_1)^3 D(-\omega_1)} \quad , \quad (4f)$$

$$\chi^{(3)}(\omega_1;-\omega_1,0(\omega_1,-\omega_1)) = \left[\frac{\partial^3 V(x)}{\partial x^3}\right]_0^2 \frac{|E_1|^2 e^4 N/m^3}{\omega_0^2 D(\omega_1)^3 D(-\omega_1)} \quad , \quad (4g)$$

$$\chi^{(3)}(\omega_1;-\omega_1,2\omega_1) = \frac{1}{2}\left[\frac{\partial^3 V(x)}{\partial x^3}\right]_0^2 \frac{|E_1|^2 e^4 N/m^3}{D(\omega_1)^3 D(-\omega_1) D(2\omega_1)} \quad . \quad (4h)$$

We observe that both holomorphic contributions, which obey Scandolo's theorem [26,29], and nonholomorphic contributions are present. Among the holomorphic contributions, the term (4b) gives the usual correction to the linear resonance (dynamical Stark effect), the term (4c) originates from optical rectification and is usually neglected, while the contributions (4d) and (4e) include the two-photon absorption and the stimulated Raman scattering. We notice that the holomorphic terms (4b) and (4c) together constitute the main contribution $\bar{\chi}^{(3)}(\omega_1)$, holomorphic in the upper complex plane, as described in the Appendix A. The above described contributions have been already given in the literature [2,30] and are discussed in ref. [30], where also the problems due to the meromorphic terms (4f), (4g) and (4h) are enphasized. We observe that the nonholomorphic terms are due to the nonlinear effects of the only probe beam, (4f) corresponding to frequencies mixing, (4g) to optical rectification and (4h) to the two-photon absorption of the probe beam.

Since the last three terms are nonholomorphic in the upper complex $\omega_1$ plane their poles have to be explicitly considered in order to derive sum rules and K.K. relations for the total $\chi^{(3)}(\omega_1)$.

We notice that two types of contributions appear in the holomorphic terms and in those which are not holomorphic, one proportional to $\left[\partial^4 V(x)/\partial^4 x\right]_0$, the other to $\left[\partial^3 V(x)/\partial^3 x\right]_0^2$. The first is due to the sum of three frequencies, the second to the sum of two frequencies, one of which is already a combination of two. These are the only way possible ways to obtain a third order susceptibility $\chi^{(3)}(\omega_1)$. Obviously, if the system has particular symmetries, some terms can be equal to zero; for instance in the case of inversion symmetry all the odd derivatives of the potential are zero so that only terms (4b) and (4f) survive.

We can notice that the only poles of the upper complex plane appearing in the last three terms (4f), (4g) and (4h) correspond to the solutions of $D(-\omega_1) = 0$, i.e. :

$$\omega_1 = \pm c + id = \pm\left(\omega_0^2 - \gamma^2/4\right)^{1/2} + i\gamma/2 \qquad (5)$$

We consider the above poles in order to obtain sum rules and K.K. relations of general validity for $\chi^{(3)}(\omega_1)$. For convenience we present in fig.1 and fig.2 a visual description of the terms proportional to $|E_2|^2$ and $|E_1|^2$ respectively.

## III. SUM RULES

The general purpose of sum rules is to find the values of the moments of the real and of the imaginary part of the susceptibility. This is possible up to a given moment, above which the integral diverges. Before deriving the sum rules by direct calculations we want to show how it is possible to find linear relations between the four sum rules referred to the first two moments of the third order susceptibility $\chi^{(3)}(\omega_1)$ just by considering the following fundamental property of all susceptibility functions:

$$\chi^{(n)}(-\omega^*) = \left(\chi^{(n)}(\omega)\right)^* \qquad (6)$$

This property requires that if $\chi^{(n)}(\omega)$ has a pole in $\alpha$, it must have another pole in $-\alpha^*$, and that $\text{Re}(\chi^{(n)}(\omega))$ is even while $\text{Im}(\chi^{(n)}(\omega))$ is odd with respect to ω if ω belongs to the real space. So we obtain for the principal parts of the integrals on the real ω axis:

$$P \int_{-\infty}^{+\infty} d\omega \omega^{2m} \chi^{(n)}(\omega) = 2 \ P \int_{0}^{+\infty} d\omega \omega^{2m} \text{Re}(\chi^{(n)}(\omega)) \qquad , \qquad (7a)$$

$$P \int_{-\infty}^{\infty} d\omega \omega^{2m+1} \chi^{(n)}(\omega) = 2i \ P \int_{0}^{+\infty} d\omega \omega^{2m+1} \text{Im}(\chi^{(n)}(\omega)) \qquad , \qquad (7b)$$

for any integer m allowed by the convergence of the integrals.

Now we restrict our analisys to the case n=3, where some conclusions can be easily obtained.

Considering our expression (3c), we observe that $D(-\omega^*) = D(\omega)^*$ since it is proportional to the inverse of a first order susceptibility (3b). As a consequence we find that the two residues of the poles (5) are one the opposite of the complex of the other. We denote the values of the two residues as a+ib and -a+ib , corresponding respectively to the poles at c+id and -c+id. We define $S_0$ and $S_2$ respectively the zeroth and the second moment of the real part, while $S_1$ and $S_3$ the first and the third moment of the imaginary part of the $\chi^{(3)}_{n.h.}(\omega_1)$, i.e. the nonholomorphic contribution to the total susceptibility $\chi^{(3)}(\omega_1)$. Taking into account that the only the non-zero value of the moments of the holomorphic contributions is the third moment of the imaginary part [26], when we consider $S_0$, $S_1$ and $S_2$ we can write $\chi^{(3)}(\omega_1)$ instead of $\chi^{(3)}_{n.h.}(\omega_1)$ without any difference. We then obtain:

$$P \int_{-\infty}^{\infty} d\omega_1 \chi^{(3)}(\omega_1) = 2 \ P \int_{0}^{+\infty} \text{Re}(\chi^{(3)}(\omega)) = 2S_0 = 2\pi i(a+ib-a+ib) = -4\pi b \qquad , \qquad (8)$$

$$P \int_{-\infty}^{\infty} d\omega_1 \omega_1 \chi^{(3)}(\omega_1) = 2i \ P \int_{0}^{+\infty} d\omega_1 \omega_1 \text{Im}(\chi^{(3)}(\omega_1)) = 2iS_1 =$$
$$= 2\pi i(a+ib)(c+id) + 2\pi i(-a+ib)(-c+id) = 4\pi i(ac-bd) \qquad . \qquad (9)$$

We have found how to express the real and imaginary part of the values of the residues of the susceptibility function as linear combinations of $S_0$ and $S_1$. Since $S_2$ and $S_3$ are functions of the residues values and of the poles only, it is clear that it is possible to obtain for them an expression which depends only on the two lowest moments. The explicit calculation for $S_2$ gives:

$$P\int_{-\infty}^{\infty}d\omega_1\omega_1^2\chi^{(3)}(\omega_1) = 2\ P\int_0^{+\infty}d\omega_1\omega_1^2\text{Re}(\chi^{(3)}(\omega_1)) = 2S_2 = 2\pi i(a+ib)(c+id)^2 +$$
$$+ 2\pi i(-a+ib)(-c+id)^2 = -4\pi(2acd + b(c^2 - d^2))$$
(10)

Substituting the values of a and b as obtained in (8) and (9) we obtain:

$$S_2 = S_0(c^2 + d^2) - 2dS_1 \tag{11}$$

which, using the expression (5) for c and d, becomes:

$$S_2 = \omega_0^2 S_0 - \gamma S_1 \tag{12}$$

Performing a similar calculation for the third moment of the imaginary part we obtain for the non-holomorphic contribution $\chi_{n.h.}^{(3)}(\omega_1)$:

$$P\int_{-\infty}^{\infty}d\omega_1\omega_1^3\chi_{n.h.}^{(3)}(\omega_1) = 2i\ P\int_0^{+\infty}\omega_1^3\text{Im}(\chi_{n.h.}^{(3)}(\omega_1)) = 2iS_3 = 2\pi i(a+ib)(c+id)^3 +$$
$$+ 2\pi i(-a+ib)(-c+id)^3 = 4\pi i(ac(c^2 - 3d^2) - bd(3c^2 - d^2))$$
(13)

Performing the same substitutions as above we obtain:

$$S_3 = 2S_0 d(c^2 + d^2) + S_1(c^2 - 3d^2) \tag{14}$$

Using expression (5) we have:

$$S_3 = S_0\gamma\omega_0^2 + S_1(\omega_0^2 - \gamma^2) \tag{15}$$

Having established the above general results we proceed to the direct calculations of the moments of the third order susceptibility. We observe that we can express the sum rules as

$$P\int_0^{\infty}d\omega_1\omega_1^{2m+1}\text{Im}(\chi^{(3)}(\omega_1)) \tag{16}$$

and

$$P\int_0^\infty d\omega_1 \omega_1^{2m} \operatorname{Re}(\chi^{(3)}(\omega_1)) \tag{17}$$

with m=0,1 since higher moments of the susceptibility would diverge because the terms (4b) and (4c) asymptotically decrease as $\omega_1^{-4}$. We perform the integrals of the terms (4f), (4g) and (4h) closing the contour in the upper complex $\omega_1$ half-plane and summing all the residues of the poles inside the semicircle considered, while we use the K.K. relations and the superconvergence theorem to compute the contributions of the holomorphic terms [26,27], obtaining immediately that all the moments of the terms (4d) and (4e) are zero because of their fast asymptotic decrease ($\omega_1^{-6}$). After lenghty calculations we finally obtain the general sum rules of the total susceptibility $\chi^{(3)}(\omega_1)$

$$P\int_0^\infty d\omega_1 \operatorname{Re}(\chi^{(3)}(\omega_1)) = -\frac{\pi|E_1|^2(\gamma^2-\omega_0^2)e^4\,N/m^3}{16\gamma^3\omega_0^6}\left[\frac{\partial^4 V(x)}{\partial x^4}\right]_0 +$$
$$+\frac{\pi|E_1|^2(14\gamma^4-11\gamma^2\omega_0^2-5\omega_0^4)e^4\,N/m^3}{48\gamma^3\omega_0^8(2\gamma^2+\omega_0^2)}\left[\frac{\partial^3 V(x)}{\partial x^3}\right]_0^2 \tag{18}$$

$$P\int_0^\infty d\omega_1 \omega_1^2 \operatorname{Re}(\chi^{(3)}(\omega_1)) = \frac{\pi|E_1|^2 e^4\,N/m^3}{16\gamma^3\omega_0^2}\left[\frac{\partial^4 V(x)}{\partial x^4}\right]_0 +$$
$$-\frac{\pi|E_1|^2(14\gamma^2+5\omega_0^2)e^4\,N/m^3}{48\gamma^3\omega_0^4(2\gamma^2+\omega_0^2)}\left[\frac{\partial^3 V(x)}{\partial x^3}\right]_0^2 \tag{19}$$

$$P\int_0^\infty d\omega_1 \omega_1 \operatorname{Im}(\chi^{(3)}(\omega_1)) = -\frac{\pi|E_1|^2 e^4\,N/m^3}{16\gamma^2\omega_0^4}\left[\frac{\partial^4 V(x)}{\partial x^4}\right]_0 +$$
$$+\frac{\pi|E_1|^2(14\gamma^2+3\omega_0^2)e^4\,N/m^3}{48\gamma^2\omega_0^6(2\gamma^2+\omega_0^2)}\left[\frac{\partial^3 V(x)}{\partial x^3}\right]_0^2 \tag{20}$$

$$P\int_0^\infty d\omega_1 \omega_1^3 \operatorname{Im}(\chi^{(3)}(\omega_1)) = \frac{\pi|E_2|^2 e^4\,N/m^3}{2(\gamma^2\omega_2^2+(\omega_2^2-\omega_0^2))}\left[\frac{\partial^4 V(x)}{\partial x^4}\right]_0 +$$

$$-\frac{\pi|E_2|^2 e^4 N/m^3}{2\omega_0^2(\gamma^2\omega_2^2+(\omega_2^2-\omega_0^2))}\left[\frac{\partial^3 V(x)}{\partial x^3}\right]_0^2 - \frac{\pi|E_1|^2 e^4 N/m^3}{24\gamma^2\omega_0^2(2\gamma^2+\omega_0^2)}\left[\frac{\partial^3 V(x)}{\partial x^3}\right]_0^2 \qquad (21)$$

We underline that the only parts proportional to the intensity of the pump beam are the first two terms of the third momentum of the imaginary part, which originate from the holomorphic contribution $\bar{\chi}^{(3)}(\omega_1)$. They correspond to those previously derived [27]. However another term, proportional to the probe intensity, appears in this moment. This third term is $S_3$, while we can identify (18) as $S_0$, (19) as $S_2$, (20) as $S_1$. So all the $S_i$ with i =0,1,2,3 are proportional to the probe intensity. We have also the confirmation that all these terms obey the linear relations (12) and (15), as can easily be checked with algebraic calculations.

## IV. KRAMERS-KRÖNIG RELATIONS

It is useful to verify the existence of specific K.K. relations and to show how they are modified with respect to those already obtained for the holomorphic contribution $\bar{\chi}^{(3)}(\omega_1)$ [26].

The method is to consider explicitely the poles in the upper complex plane and to perform the Cauchy integrals for all the moments which are allowed by the asympotic behaviour. The asymptotic behaviour of $\chi^{(3)}(\omega_1)$ is $\propto \omega_1^{-4}$, so that the moments to be considered are $\chi^{(3)}(\omega_1)$ and $\omega_1^2 \chi^{(3)}(\omega_1)$. We write explicitly the two Cauchy integrals, considering the poles at (c+id) and at (-c+id). We obtain, taking into account the relations between the residues and the results of the sum rules shown in the previous section:

$$P\int_{-\infty}^{+\infty}d\omega_1 \frac{\chi^{(3)}(\omega_1)}{\omega_1-\omega} = \pi i \chi^{(3)}(\omega) + 2\pi i\left(\frac{a+ib}{c+id-\omega}+\frac{-a+ib}{-c+id-\omega}\right)=$$
$$= \pi i \chi^{(3)}(\omega) + \frac{2(\omega-i\gamma)S_0}{D(-\omega)} + \frac{2iS_1}{D(-\omega)} \qquad (22)$$

and

$$P\int_{-\infty}^{+\infty}d\omega_1\frac{\omega_1^2\chi^{(3)}(\omega_1)}{\omega_1-\omega}=\pi i\omega^2\chi^{(3)}(\omega)+2\pi i\left(\frac{(a+ib)(c+id)^2}{c+id-\omega}+\frac{(a+ib)(-c+id)^2}{-c+id-\omega}\right)$$

$$=\pi i\omega^2\chi^{(3)}(\omega)+\frac{2(\omega-i\gamma)S_2}{D(-\omega)}+\frac{2iS_3}{D(-\omega)}=\pi i v^2\chi^{(3)}(\omega)+\frac{2\omega\omega_0^2 S_0}{D(-\omega)}+\frac{2i(\omega_0^2+i\gamma\omega)S_1}{D(-\omega)} \qquad (23)$$

We can separate real and imaginary part of the preceding expressions and obtain two couples of K.K. relations in the usual form, the first one originating from (22):

$$\begin{cases}\text{Re}(\chi^{(3)}(\omega))=\dfrac{2}{\pi}P\int_0^{+\infty}d\omega_1\dfrac{\omega_1\,\text{Im}(\chi^{(3)}(\omega_1))}{\omega_1^2-\omega^2}-\dfrac{2\omega}{\pi}S_0\,\text{Im}\left(\dfrac{1}{D(-\omega)}\right)-\dfrac{2}{\pi}(S_1-\gamma S_0)\text{Re}\left(\dfrac{1}{D(-\omega)}\right)\\[2mm]\text{Im}(\chi^{(3)}(\omega))=-\dfrac{2\omega}{\pi}P\int_0^{+\infty}d\omega_1\dfrac{\text{Re}(\chi^{(3)}(\omega_1))}{\omega_1^2-\omega^2}+\dfrac{2\omega}{\pi}S_0\,\text{Re}\left(\dfrac{1}{D(-\omega)}\right)-\dfrac{2}{\pi}(S_1-\gamma S_0)\text{Im}\left(\dfrac{1}{D(-\omega)}\right)\end{cases}, \quad (24)$$

and the second one originating from (23):

$$\begin{cases}\omega^2\text{Re}(\chi^{(3)}(\omega))=\dfrac{2}{\pi}P\int_0^{+\infty}d\omega_1\dfrac{\omega_1^3\,\text{Im}(\chi^{(3)}(\omega_1))}{\omega_1^2-\omega^2}-\dfrac{2\omega}{\pi}S_2\,\text{Im}\left(\dfrac{1}{D(-\omega)}\right)+\dfrac{2}{\pi}(S_3-\gamma S_2)\text{Re}\left(\dfrac{1}{D(-\omega)}\right)\\[2mm]\omega^2\text{Im}(\chi^{(3)}(\omega))=-\dfrac{2\omega}{\pi}P\int_0^{+\infty}d\omega_1\dfrac{\omega_1^2\,\text{Re}(\chi^{(3)}(\omega_1))}{\omega_1^2-\omega^2}+\dfrac{2\omega}{\pi}S_2\,\text{Re}\left(\dfrac{1}{D(-\omega)}\right)-\dfrac{2}{\pi}(S_3-\gamma S_2)\text{Im}\left(\dfrac{1}{D(-\omega)}\right)\end{cases}. \quad (25)$$

Explicitating $S_2$ and $S_3$ with the help of expressions (12) and (15) we obtain the following K.K. expression in terms of the lowest moments:

$$\begin{cases}\omega^2\text{Re}(\chi^{(3)}(\omega))=\dfrac{2}{\pi}P\int_0^{+\infty}d\omega_1\dfrac{\omega_1^3\,\text{Im}(\chi^{(3)}(\omega_1))}{\omega_1^2-\omega^2}-\dfrac{2\omega}{\pi}(\omega_0^2 S_0-\gamma S_1)\text{Im}\left(\dfrac{1}{D(-\omega)}\right)+\dfrac{2}{\pi}\omega_0^2 S_1\,\text{Re}\left(\dfrac{1}{D(-\omega)}\right)\\[2mm]\omega^2\text{Im}(\chi^{(3)}(\omega))=-\dfrac{2\omega}{\pi}P\int_0^{+\infty}d\omega_1\dfrac{\omega_1^2\,\text{Re}(\chi^{(3)}(\omega_1))}{\omega_1^2-\omega^2}+\dfrac{2\omega}{\pi}(\omega_0^2 S_0-\gamma S_1)\text{Re}\left(\dfrac{1}{D(-\omega)}\right)-\dfrac{2}{\pi}\omega_0^2 S_1\,\text{Im}\left(\dfrac{1}{D(-\omega)}\right)\end{cases}. \quad (26)$$

It is interesting to note the perfect simmetry between the two K.K. relations (24) and (25) with respect to the terms originating from non-holomorphic contributions: it can be observed a clear correspondence between $S_0$ and $S_1$ in the (24) with respectively $S_2$ and $S_3$ in the (25). We can observe that all the terms originating from non-holomorphic contributions are linear with respect to the $S_i$ with i=0,1,2,3.

The nonlinear K.K. relations we have here obtained are more general than the ones given before, and in each of them we have also terms which are proportional to the intensity of the probe beam and are directly related to the two lowest moments of the susceptibility $S_0$ and $S_1$. Our procedure is a direct one, and avoids the use of the maximum entropy hypothesis [37,38] adopted by Peiponen [30,39,40,41].

Our K.K. analyses, as given by (24) and (25), are analytically complete and overcome the difficulties explicitely shown in ref. [30] (see for instance fig. 3.6 at p.60 and fig. 5.5 at p. 92). In comparison with the results of the maximum entropy method shown in fig. 5.5 at p. 92 in ref. [30], our results reproduce exactly the relations between the real and the imaginary part of the susceptibility, since no hypotheses or approximations have been made.

## V. CONCLUSIONS

We wish to summarize the main results obtained above as follows.
The anharmonic oscillator model has been used to derive general expressions for the nonlinear susceptibility with Pump and Probe optical beams. The lowest order susceptibility $\chi^{(3)}(\omega_1)$ is shown to contain terms proportional to $|E_2|^2$ and also terms proportional to $|E_1|^2$. The former are holomorphic in the upper complex plane with respect to the variable $\omega_1$ and consequently obey sum rules and Kramers-Krönig relations of the type derived previously. The latter contain poles in the upper complex $\omega_1$ plane, and give contributions to the nonlinear sum rules and K.K. relations which had been formerly neglected. These contributions are shown to depend on the zero and first moment of the susceptibility only.

The above contributions to the sum rules have been shown to depend on the derivatives of the potential at the equilibrium position, the third order susceptibility depending on the fourth order derivative and on the square of the third order derivative.

The above results are expected to be of general significance, in spite of the simplicity of the model adopted, because they are only determined by the analytic properties of the response function due to time causality. It can be shown in fact that the leading holomorphic term here considered, $\chi^{(3)}(\omega_1;\omega_1,-\omega_2,\omega_2)$, gives the same result in the most general physical system, treated with the Kubo response function [42], provided one substitutes the expectation values of the derivatives for the values at the equilibrium position., and one considers the corresponding tensor which obtains from the spatial directions of the applied fields.

The results here presented can be of interest for a detailed analysis of experimental data because they show explicitely the dependence on the intensity of the probe beam itself which must appear in the third order susceptibility $\chi^{(3)}(\omega_1)$. Experimentally it is also possible to separate the contribution proportional to the pump intensity by using modulation techniques so that the holomorphic and the nonholomorphic contributions can be separately measured. We suggest experiments where both the probe frequency and the probe intensity can be varied, so as to verify the appearence of the additional contributions in the sum rules and K.K. relations.

## ACKNOLEDGMENTS

We are grateful to S. Scandolo and to G. La Rocca for suggestions and useful discussions.

## APPENDIX A

In this appendix we present the general formula for the $n^{th}$ order Polarization at the generic frequency $r\omega_1+s\omega_2$, as derived from the iterative procedure described in ref.[33]:

$$P^{(n)}(r\omega_1 + s\omega_2) = -\frac{1}{D(r\omega_1 + s\omega_2)} \sum_{\substack{i,k,l \\ t,p,q}} \left( \frac{1}{e^{\left(\sum_{j=1}^{l}\sum_{h=1}^{t_j} k_{j,h}-1\right)} N^{\left(\sum_{j=1}^{l}\sum_{h=1}^{t_j} k_{j,h}-1\right)}} \times \right.$$

$$\left. \times \left[ \frac{\partial^{\left(\sum_{j=1}^{l}\sum_{h=1}^{t_j} k_{j,h}+1\right)} V(x)}{\partial x^{\left(\sum_{j=1}^{l}\sum_{h=1}^{t_j} k_{j,h}+1\right)}} \right]_0 \prod_{j=1}^{l}\prod_{h=1}^{t_j} \frac{\left(P^{(i_{j,h})}(p_{j,h}\omega_1 + q_{j,h}\omega_2)\right)^{k_{j,h}}}{k_{j,h}!} \right) \quad (A1a)$$

with the conditions:

$$1 \leq j \leq l \quad , 1 \leq h, h' \leq t_j$$

$$\left(1 \leq i_{j,h} < i_{j+1,h'} < n\right) \text{ and } \left(h \neq h' \Rightarrow \left(p_{j,h} \neq p_{j,h'}\right) \quad \text{or} \quad \left(q_{j,h} \neq q_{j,h'}\right)\right) \quad . \quad \text{(A1b)}$$

The frequency considered ($r\omega_1+s\omega_2$) imposes the following constraint:

$$\sum_{j=1}^{l}\sum_{h=1}^{t_j}\left(p_{j,h}\omega_1 + q_{j,h}\omega_2\right)k_{j,h} = r\omega_1 + s\omega_2 \quad ; \quad \text{(A1c)}$$

and the order to which Polarization is considered gives:

$$\sum_{j=1}^{l}\sum_{h=1}^{t_j} k_{j,h} i_{j,h} = n \quad . \quad \text{(A1d)}$$

In order to consider Pump and Probe experiments we specialize the expression for the probe frequency imposing r=1 and s=0 in expression (4) and define the following quantity, which immediately results to be limited to odd n :

$$\chi^{(n)}(\omega_1) \equiv \frac{P^{(n)}(\omega_1)}{E_1} \quad . \quad \text{(A2)}$$

The above expressions contain terms which are predominant for an intense pump beam and a weak probe, and are holomorphic in the upper complex $\omega_1$ plane. This corresponds to the contribution considered in the Kubo [42] expression and used in [26] to derive K.K. relations. Such terms are:

$$\overline{\chi}^{(n)}(\omega_1) = -\frac{P^{(1)}(\omega_1)}{E_1 D(\omega_1)} \sum_{\substack{i,k,l \\ t,q}} \left( \frac{1}{e^{\left(\sum_{j=1}^{l}\sum_{h=1}^{t_j} k_{j,h}\right)} N^{\left(\sum_{j=1}^{l}\sum_{h=1}^{t_j} k_{j,h}\right)}} \times \right.$$

$$\times \left[ \frac{\partial^{\left(\sum_{j=1}^{l}\sum_{h=1}^{t_j}k_{j,h}+2\right)} V(x)}{\partial x^{\left(\sum_{j=1}^{l}\sum_{h=1}^{t_j}k_{j,h}+2\right)}} \right]_0 \prod_{j=1}^{l}\prod_{h=1}^{t_j} \frac{\left(P^{(i_{j,h})}(q_{j,h}\omega_2)\right)^{k_{j,h}}}{k_{j,h}!} \quad , \tag{A3a}$$

with the conditions:

$$1 \leq j \leq l \quad , 1 \leq h, h' \leq t_j$$

$$\left(1 \leq i_{j,h} < i_{j+1,h'} < n\right) \text{ and } \left(h \neq h' \Rightarrow \left(q_{j,h} \neq q_{j,h'}\right)\right) \quad , \tag{A3b}$$

and the usual constraints regarding rispectively order of Polarization and frequency :

$$\sum_{j=1}^{l}\sum_{h=1}^{t_j} k_{j,h} i_{j,h} = n-1 \quad ; \tag{A3c}$$

$$\sum_{j=1}^{l}\sum_{h=1}^{t_j} q_{j,h} k_{j,h} = 0 \quad ; \tag{A3d}$$

with an additional condition to select those terms in the summation (6a) which are not functions of $\omega_1$:

$$\frac{\partial\left(P^{(i_{j,s})}(\beta_{j,s}\omega_2)\right)}{\partial \omega_1} = 0 \quad . \tag{A3e}$$

The main properties of above described terms are the following:
i) they are proportional to the intensity of the pump beam to the highest possible power, the $(n-1)/2$ th;
ii) they are olomorphic with respect to the variable $\omega_1$ in the upper complex half-plane;
iii) they have the slowest possible decrease at infinity with respect to $\omega_1$ ($\propto \omega_1^{-4}$)

The first and third properties give them the predominance over the other terms of the same order; the second and the third one allow us to write two K.K. relations because of Titmarsch's theorem, one for $\overline{\chi}^{(n)}(\omega_1)$ and the second one for $\omega_1^2 \overline{\chi}^{(n)}(\omega_1)$, as shown in ref. [26]. It is also possible to prove that the summation in (6a) for every n results to be real by using the relation $P^{(m)}(-v^*) = \left(P^{(m)}(v)\right)^*$, which holds for every m, and rearranging the terms in thr summation so that conjugate terms appear in couple.

The preceding considerations and the analysis of the asymptotic behaviour as obtained from K.K. relations and from expression (A3) allow us to derive sum rules for the moments of the $\overline{\chi}^{(n)}(\omega_1)$ up to the third moment of the imaginary part, which is the only one different from zero. Here we give its explicit expression:

$$\int_0^\infty d\omega_1 \omega_1^3 \, \text{Im}\overline{\chi}^{(n)}(\omega_1) = \frac{\pi}{2E_1} \sum_{\substack{i,k,l \\ t,q}} \left( \frac{1}{e^{\left(\sum_{j=1}^{l}\sum_{h=1}^{t_j} k_{j,h}\right)} N^{\left(\sum_{j=1}^{l}\sum_{h=1}^{t_j} k_{j,h}\right)}} \times \left[ \frac{\partial^{\left(\sum_{j=1}^{l}\sum_{h=1}^{t_j} k_{j,h}+2\right)} V(x)}{\partial x^{\left(\sum_{j=1}^{l}\sum_{h=1}^{t_j} k_{j,h}+2\right)}} \right]_0 \prod_{j=1}^{l}\prod_{h=1}^{t_j} \frac{\left(P^{(i_{j,h})}(q_{j,h}\omega_2)\right)^{k_{j,h}}}{k_{j,h}!} \right) \quad (A4)$$

with the same constraints (A3b-A3c-A3d-A3e) in the summations. It can easily verified that the above results are in agreement with the general results obtained previously for the third order susceptibility [26].